\documentclass[apjl]{emulateapj}

\shorttitle{Two new brown dwarf binaries}
\shortauthors{Artigau et al.}

\begin{document}

\title{ Discovery of two L \& T binaries with wide separations and peculiar photometric properties }

\author{\'Etienne Artigau\altaffilmark{1}, David Lafreni\`ere\altaffilmark{1}, Ren\'e Doyon\altaffilmark{1}, Michael Liu\altaffilmark{2,3}, Trent J. Dupuy\altaffilmark{2}, Lo\"{\i}c Albert\altaffilmark{1,4}, Jonathan Gagn\'e\altaffilmark{1}, Lison Malo\altaffilmark{1},  Damien Gratadour\altaffilmark{5}}

\newcommand{\ang}{\AA\hspace{.01cc} } 
\newcommand{\simpbint}{SIMP~J1619+0313}
\newcommand{\simpbinl}{SIMP~J1501-0135}
\newcommand{\Ks}{\mbox{$K_{\rm s}$}}
\newcommand{\degs}{\mbox{$^{\circ}$}}
\newcommand{\etal}{et al.}
\newcommand{\eg}{e.g.}
\newcommand{\ie}{i.e.}

\altaffiltext{1}{D\'epartement de Physique and Observatoire du Mont-M\'egantic, Universit\'e de Montr\'eal, C.P. 6128, Succ. Centre-Ville, Montr\'eal, QC, H3C 3J7, Canada}
\altaffiltext{2}{Institute for Astronomy, University of Hawai`i, 2680 Woodlawn Drive, Honolulu, HI 96822}
\altaffiltext{3}{Alfred P. Sloan Research Fellow; mliu@ifa.hawaii.edu}
\altaffiltext{4}{Canada-France-Hawaii Telescope Corporation, 65-1238 Mamalahoa Highway, Kamuela, HI 96743, USA}
\altaffiltext{5}{Laboratoire d'\'Etudes Spatiales et d'Instrumentation en Astrophysique (LESIA), Observatoire de Paris, CNRS, UPMC, Universit\'e Paris Diderot, 5 places Jules Janssen, 92195 Meudon France}

\email{artigau@astro.umontreal.ca}

\begin{abstract}

We present spatially resolved photometric and spectroscopic observations of two wide brown dwarf binaries uncovered by the SIMP near-infrared proper motion survey. The first pair  (SIMP\,J1619275+031350AB) has a separation of $0.691\arcsec$ (15.2 AU) and components T2.5+T4.0, at the cooler end of the ill-understood $J$-band brightening. The system is unusual in that the earlier-type primary is bluer in $J-K_{\rm s}$ than the later-type secondary, whereas the reverse is expected for binaries in the late-L to T dwarf range. This remarkable color reversal can possibly be explained by very different cloud properties between the two components. The second pair (SIMP\,J1501530-013506AB) consists of an L4.5+L5.5 (separation $0.96\arcsec$, 30-47\,AU) with a surprisingly large flux ratio ($\Delta J =1.79$\,mag) considering the similar spectral types of its components. The large flux ratio could be explained if the primary is itself an equal-luminosity binary, which would make it one of the first known triple brown dwarf systems. Adaptive optics observations could not confirm this hypothesis, but it remains a likely one, which may be verified by high-resolution near-infrared spectroscopy. These two systems add to the handful of known brown dwarf binaries amenable to resolved spectroscopy without the aid of adaptive optics and constitute prime targets to test brown dwarf atmosphere models.

\end{abstract}

\section{Introduction}

Binarity is relatively common among brown dwarfs (BDs) and recent work on these objects has shown that their properties differ from those of stellar binaries on many levels (see the review by \citealt{Burgasser2007} and references therein). Their binarity fraction (10-30\%) is lower than for Solar-type stars ($\sim65\%$, \citealt{Duquennoy1991}) and follows the overall trend of decreasing binarity fraction with decreasing mass. Field brown dwarf binary systems are tighter than their more massive cousins, with few known systems with separations beyond 20\,AU ($\textless$2.3\% of M7-L8 dwarfs have wide, 40 to 1000 AU, companions; \citealt{Allen2007a}). Their mass ratios are also closer to unity (77\% of systems have a mass ratio $\ge 0.8$; \citealt{Burgasser2007}). These different properties contain valuable information regarding their mechanism of formation. The dearth of wide BD binaries might be a signature of relatively violent formation processes: accreting proto-stellar cores that are ejected from the cluster before reaching a stellar mass \citep{Reipurth2001} or photo-evaporation of pre-stellar cores by nearby massive stars \citep{Whitworth2004}. Alternative processes have been proposed: gravitational collapse within a circumstellar disk \citep{Boss2000} and collapse of very low mass molecular cloud cores \citep{Boyd2005, Padoan2004, Padoan2005}. Observations of disks \citep{Liu2003} and outflows around young BDs suggest that they undergo a T-Tauri-like phase, which points to a star-like formation mechanism \citep{Jayawardhana2003, Mohanty2005, Scholz2006}. Each of these formation mechanisms leads to different BD binary frequency, separation distribution, and mass ratio distribution.

Beyond the interest in constraining BD formation models, binary BDs are natural benchmark objects to test atmosphere and evolutionary models. Their equal metallicity, coevality, and common distance put strong constraints when estimating these parameters from models by fitting both objects simultaneously \citep{Liu2010}. Binaries that challenge evolutionary models have been found, most notably the 2MASS J05352184-0546085 system discovered by \citet{Stassun2006} that shows a temperature reversal (the least massive object being the warmer one) that has tentatively been explained by widely different magnetic activity in the two components \citep{Reiners2007}.

Binary BDs overlapping with the $J$-band brightening (roughly late-L to mid-T) are of particular interest as theoretical models cannot easily reproduce the observed evolution of BDs in this interval. The $J$-band brightening is partially explained by the presence of increased binarity at the L/T transition \citep{Liu2006}. This is particularly the case for objects with the strongest brightening. However, even taking into account confirmed and suspected binaries, a smaller, but still significant, brightening is observed. This is also confirmed by the existence of binaries with flux reversals \citep{Liu2006, Looper2008}. This brightening is probably best understood as resulting from the disappearance of dust-bearing clouds from the photosphere, but this temperature interval has proved challenging to describe with a self-consistent atmosphere model (\citealt{Marley2010} and references therein). Binaries, especially those amenable to spatially resolved spectroscopy, provide unique constraints to these models. Tighter binaries have orbital periods short enough for orbital monitoring and provide further constraints on models through dynamical masses \citep{Liu2008, Dupuy2009, Konopacky2010}.

In 2006, we have undertaken a near-infrared proper motion survey ({\it Sondage Infrarouge de Mouvement Propre} - SIMP; \citealt{Artigau2009}) with the wide-field near-infrared camera CPAPIR \citep{Artigau2004} at the CTIO 1.5\,m and OMM 1.6\,m \citep{Racine1978} telescopes and have covered $\sim35\%$ of the sky up to now. BD candidates were found using both the SIMP and 2MASS database obtained at different epochs to identify high proper motion sources. Spectroscopic follow-up of high proper motion candidates has been done with GNIRS \citep{Elias2006}, SpeX \citep{Rayner2003} and NIRI \citep{Hodapp2003} in 2006, 2007 and 2008. More than 80 new L dwarfs and 14 new T dwarfs (Artigau et al., in prep.) were confirmed. 

We did not expect to detect any resolved binaries in our sample given that the  binary BDs separation distribution peaks at $\sim4$\,AU \citep{Maxted2005, Burgasser2007} and that all of our observations were seeing-limited. We nevertheless searched systematically through our spectroscopy acquisition images and found two partially resolved binaries: a pair of mid-L dwarfs separated by $\sim1.0\arcsec$ (SIMPJ1501530-013506, hereafter {\simpbinl}) and a pair of mid-Ts separated by $\sim0.7\arcsec$ (SIMPJ1619275+031350, hereafter {\simpbint}). Following these discoveries, we obtained a series of observations to characterize the individual components of these systems.

In \textsection~\ref{spectro} we present the NIRI resolved spectroscopy of both systems. In \textsection~\ref{niri_photometry}, \textsection~\ref{megacam_photometry}, \textsection~\ref{cpapir_photometry} and \textsection~\ref{lgs} we detail resolved and unresolved, seeing-limited, $i$ through $K_{\rm s}$ photometric measurements and Laser Guide Star (LGS) AO observations of both binaries. Finally, the spectral typing of all components is discussed in \textsection~\ref{typing} and \textsection~\ref{discussion} details the properties of both systems.

\section{Observations}

Following the discovery of the binary nature of \simpbinl\ and \simpbint\, we obtained a set of observations to characterize their resolved far-red and near-infrared  (0.7-2.45\,$\mu$m) spectral energy distribution (SED). These observations were aimed at putting both systems in the broader context of the existing sample of resolved BD binaries, as they both showed peculiarities worthy of further study. The large contrast ratio of \simpbinl\ ($\Delta H \sim 1.5\,$mag) suggested a mid-L/mid-T binary, of which we know only a handful of examples. {\simpbint}, with a $\sim$T3 blended spectral type, lies at the cooler end of the $J$-band brightening and was seen as likely to provide useful constraints to models attempting to reproduce this interval.

The characterization was done through resolved optical and near-infrared photometry, resolved infrared spectroscopy, high-angular resolution imaging, and accurate near-infrared blended photometry. These observations resulted in detailed portraits of the individual components of both systems. A log of all observations is provided in Table~\ref{tbl-3}. All measurements available at hand, both from archives and from the observations presented here, as well as derived quantities, are compiled for \simpbint\ and \simpbinl\ in Tables~\ref{tbl-1} and \ref{tbl-2} respectively. When applicable, these measurements are given for the individual components of both systems.

\subsection{NIRI resolved spectroscopy}
\label{spectro}

Seeing limited spectroscopy was obtained at the Gemini North telescope with NIRI for both binaries under excellent seeing conditions (${\rm FWHM}\sim0.4\arcsec$). For both binaries, the slit was aligned with the binary axis to include both components in the slit. The 4-pixel (0.409\arcsec) wide slit in the blue setup was used in combination with the f/6 $J$, f/6 $H$ and f/6 $K$ grisms for a resolving power of 650 to 825. For each grism setup, a set of 10 120-s exposures was taken with a 5\arcsec\ nod along the slit between exposures. Telluric correction was performed using A0-A1 star spectra (HIP 75230 and HIP 79463 for {\simpbint} and {\simpbinl}, respectively). Spectroscopic flats were obtained as part of the standard calibrations with the GCAL calibration unit. Wavelength calibration was performed by registering bright telluric emission lines.

\begin{figure*}[!htb]
\epsscale{.99}
\plotone{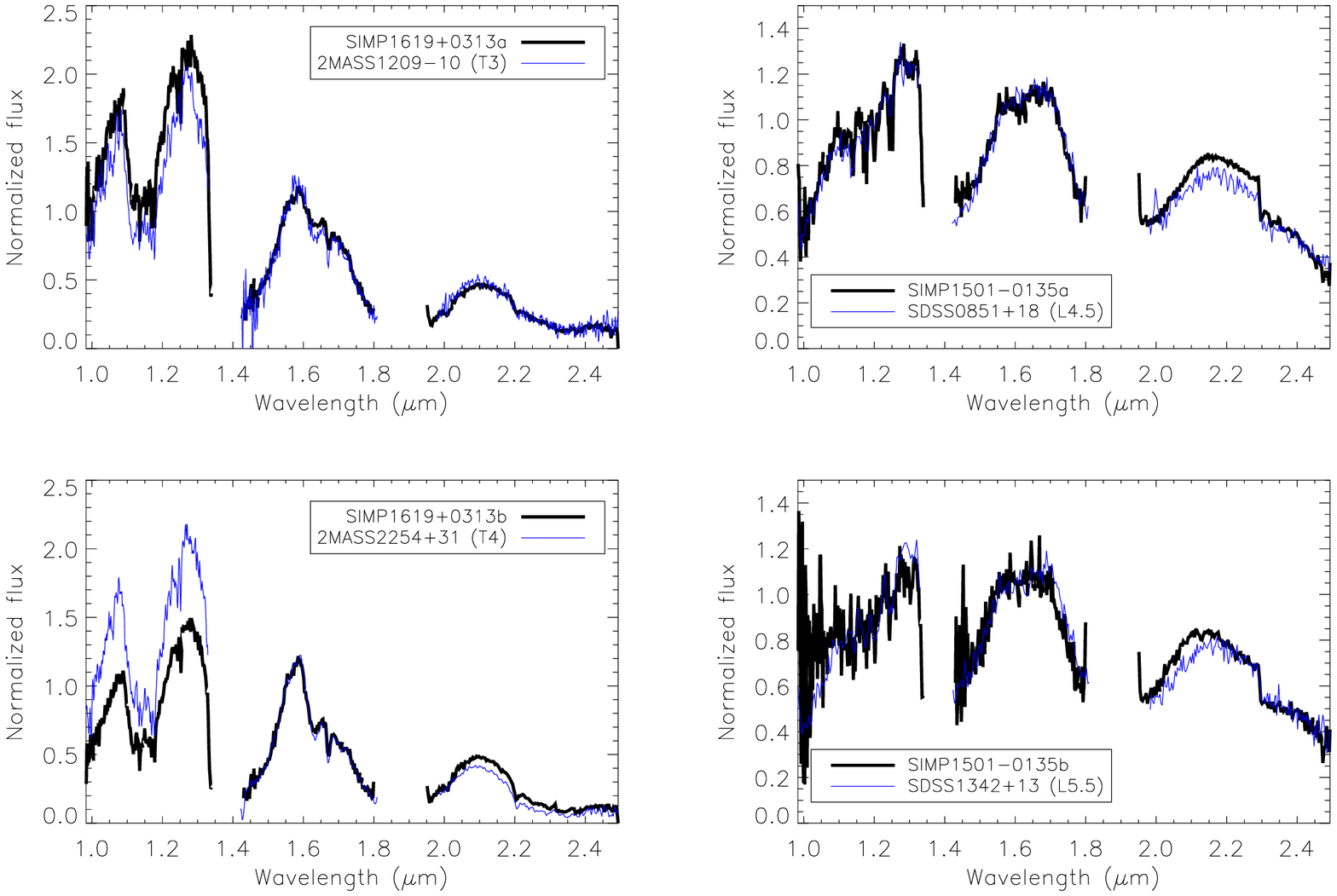}
\caption{ \label{fig3} Near-infrared spectra of {\simpbint}A (upper left), {\simpbint}B (upper right),  {\simpbinl}A (lower left) and {\simpbinl}B (lower right) taken with NIRI at Gemini North. The {\simpbinl}A and B spectra are over-plotted with the archival spectra of the L4.5 SDSS0851+18 and L5.5 SDSS1342+13, while the spectra of {\simpbint}A and B are over-plotted with these of the T3 2MASSJ 1209-10 and T4 2MASSJ2254+31, respectively (\citealt{Chiu2006, Knapp2004}). All spectra are normalized to their median flux over the $1.55-1.60$\,$\mu$m interval.}
\end{figure*}

The spectra were reduced by first pair subtracting two dithered spectral images. The resulting positive and negative traces showed the spectra of both components with little overlap. We extracted the individual spectra by using a linear inversion method where, for each spectral pixel,  we represented the trace along the spatial direction by the sum of two 1-D Gaussians of identical widths, with the same separation as the binary. This linear inversion correctly handles the overlapping spectra. Once extracted, the final spectrum of each component was obtained by taking the median of all individual exposures and corrected for telluric absorption. The $J$, $H$ and $K$ spectra were scaled by adjusting synthetic fluxes to the NIRI and CPAPIR photometry (\textsection~\ref{niri_photometry}  and \textsection~\ref{cpapir_photometry}). Figure \ref{fig3} shows the resolved spectra of both components of both binaries compared to archival spectra of L4.5, L5.5, T3 and T4 dwarfs.

\subsection{NIRI resolved photometry}
\label{niri_photometry}
In addition to the spectroscopic observations, we obtained resolved $J$, $H$, {\Ks} photometry with NIRI for both binaries under excellent seeing conditions (0.45-0.52\arcsec). In addition, these observations confirm the differential photometry obtained from AO observations (\textsection~\ref{lgs}). Resolved photometry is necessary to accurately reconstruct the near-infrared spectrum of each binary component as the NIRI spectra were obtained piecewise with separate $J$, $H$ and $K$-grism setups.

The images were reduced in a standard manner. A sky image was obtained by median combining all images obtained over 5 dither positions, then, all science images were sky subtracted and flat-fielded. We obtained accurate flux ratios by performing simultaneous PSF fitting on the two resolved components of the pairs for each individual science frame. The photometric accuracy was estimated from the standard deviation of the flux ratio measurements. We did not attempt to obtain absolute photometry measurements with this dataset as only a handful of field (calibration) stars are present in each $2\arcmin\times2\arcmin$ NIRI field of view. 

\subsection{MegaCam photometry}
\label{megacam_photometry}
We obtained far-red photometry for both binary systems to further constrain their physical properties, especially in view of their peculiar near-infrared photometry. The shape of the far-red ($i$ to $z$) SED of L and T dwarfs is dominated by the pressure-broadened KI resonance doublet centered at $\sim0.77$\,$\mu$m \citep{Burrows2000}. The SED in this wavelength interval varies with surface gravity, metallicity and the dust grain content of the atmosphere and is thus complementary to near-infrared spectroscopy. 

A 200-s $i$-band snapshot of each target was obtained at CFHT with MegaCam \citep{Boulade2003}.  {\simpbinl} was observed on 2006 July 20 with a seeing of $0.95\arcsec$ while \simpbint\ was observed on 2006 August 29 with a seeing of $0.68\arcsec$. {\simpbint} was further observed in $z$ band on 2009 June 1; an exposure of 100\,s was used and the seeing was $1.1\arcsec$. The photometry is reported in the AB system and was calibrated using standard stars observed during the same observing run, without color-term corrections. The {\simpbint}AB components are marginally resolved in the MegaCam imagery (i.e. separation of about one FWHM); the flux measurements for both components listed in Table\,\ref{tbl-1} have been derived through PSF fitting using a field star as a reference PSF.  The binary \simpbinl\ is unresolved in the $i$-band image.

\subsection{CPAPIR imaging and astrometry}
\label{cpapir_photometry}
While the flux ratios of the components can be obtained through high resolution imaging, absolute photometry of the pairs needs calibration using numerous field stars. The unresolved 2MASS photometry of both binaries only has a 7-9\% precision in the three near-infrared band-passes, resulting in large (i.e. $\sim15\%$) color measurement errors, poorer by an order of magnitude than the component flux ratio measurements. To improve the blended photometry of the pairs, we obtained $J$, $H$ and {\Ks} photometry at the Observatoire du Mont-M\'egantic (OMM) using the CPAPIR \citep{Artigau2006} wide-field imager on 2009 February 17, 2009 March 3, 2009 March 4 and 2009 March 9.

Near-infrared colors of brown dwarfs can potentially vary by up to $\sim0.4$\,mag depending on the photometric system used \citep{Stephens2004}. This needs to be taken into account when comparing the photometric properties of our targets to these of other field brown dwarfs. The CPAPIR filters match those of the Mauna Kea (MKO) system \citep{Simons2002, Tokunaga2002}. The unresolved photometry measurements were performed through aperture photometry (radius of one FWHM) and are reported in Tables~\ref{tbl-1} and  \ref{tbl-2}. The zero points were set by using all 2MASS stars in the Point Source Catalog within an $8\arcmin$ radius around the targets. We converted the 2MASS magnitudes into the Mauna Kea system using the \citet{Leggett2006} polynomial relations. The uncertainties on the zero points (typically $\sim1\%$) were determined from the dispersion of differences to the median divided by the square root of the number of calibration stars.

For comparison with BDs in the literature, the Mauna Kea system magnitudes determined for both binaries were converted back into the 2MASS system using the NIRI spectra and the procedure described in \citet{Stephens2004}. The corrections  derived from this procedure match those given by the  polynomial relations of \citet{Stephens2004}  to within 2\% for the $J$ and $H$ bands. Remarkably, the $K$-band correction differs by $\sim10\%$ for both {\simpbinl} components as the $K$-band portion of their spectra peaks at a slightly bluer wavelength compared to field objects of similar spectral types (see in particular {\simpbinl}B in Figure\,\ref{fig3}). Both objects are therefore brighter in the slightly bluer $K_{\rm MKO}$ compared to $K_{\rm 2MASS}$.

The CPAPIR discovery (2006) and follow-up (2009) datasets were used to measure the proper motion of both pairs. The proper motion of field stars \citep{Zacharias2005} has been taken into account and, for both binaries, we corrected the proper motion measurement for the parallax estimated from the photometric distance. These corrections are smaller than 5 mas/yr in both spatial directions and smaller than the astrometric accuracy. Tables~\ref{tbl-1} and \ref{tbl-2} report the measured proper motions. A search for common proper motion stars within a 30\arcmin\ radius was performed using the NLTT \citep{Salim2003} catalog. No object was found to have a common PM with either binary. We note that NLTT39039 lies 15.8\arcmin\ from {\simpbinl} and has a similar PM, but our analysis rules-out a common PM.

\subsection{LGS Adaptive Optics imaging}
\label{lgs}
The relatively wide physical separation of either binary (exceeded only by the SDSS1416+13AB system, \citealt{Burningham2010}) leaves room for higher-order multiplicity for both systems. The case for a higher-order multiple is further strengthened for \simpbinl\ by its large flux ratio. Furthermore, adaptive optics observations provide accurate differential photometry that can be combined with unresolved photometry for absolute flux measurements of individual components. Both targets are too faint at visible wavelengths for natural guide star AO observations and neither has a bright enough nearby star for that purpose. However, they do both have nearby guide stars that are sufficiently bright to enable LGS AO operation. 

\begin{figure}
\epsscale{.99}
\plotone{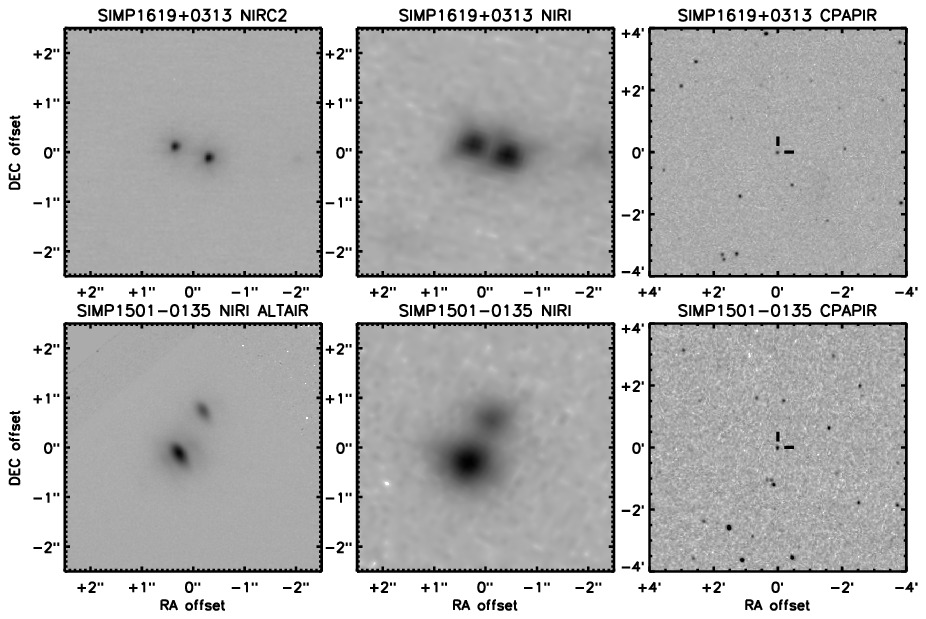}
\caption{\label{fig1} {\Ks}-band images of {\simpbinl} (top) and {\simpbint} (bottom). The right panels show images obtained with CPAPIR  as part of SIMP. The central panels show both binaries observed with NIRI at Gemini North without adaptive optics under prime observing conditions. The left panels show the follow up LGS AO images. The {\simpbinl} images shows a strong elongation due to the rather unfavorable configuration of the tip/tilt guide star which was close to the Gemini North laser system separation limit.}
\end{figure}

We obtained $J$, $H$ and {\Ks} imaging of {\simpbinl} at the Gemini North telescope using the NIRI imager in its f/32 configuration with the Altair AO system operated in LGS mode. Altair was used with its field lens that significantly reduces the anisoplanatism effects. The tip-tilt star (USNO-B1.0~0884-0256280, $R=16.5$; \citealt{Monet2003}) was located 37\arcsec\ from our target, at the workable limit of the system. We therefore had to offset {\simpbinl} from the center of the field. The images were reduced in a standard manner: a sky image was constructed from the median combination of the dithered science sequence, then science images were sky subtracted and divided by a dome flat. The images were then registered to a common center and median combined. For each filter, a set of 9 45-s images dithered over a $2.5$\arcsec\ pattern was taken. The separation and position angle of the binary were taken to be the mean of the values measured in the individual \Ks\ frames and the uncertainty was derived from the standard deviation of the individual values. Note that the NIRI Altair plate scale has not been fully characterized, we assumed a pixel scale of 21.5\,mas/pixel\footnote{The NIRI+Altair plate-scale has been better characterized without the field lens, in which case it has been measured to be 21.9\,mas/pix. Differential plate-scale measurements indicate that the introduction of the field lens reduces the plate-scale by $1.6\%$, hence the 21.5\,mas/pix scale used here.} and adopted a rather conservative 1\% uncertainty in the separation of the binary. The resulting {\Ks} image of the pair is shown in Figure\,\ref{fig1}. The PSF elongation is due to the sub-optimal LGS configuration.

\simpbint\ was observed on 2008~April~01 and May~29 using the sodium LGS AO system of the 10-meter Keck II Telescope on Mauna Kea, Hawaii \citep{vanDam2006, Wizinowich2006}. Conditions were photometric for both runs.  The facility NIRC2  IR camera was used both in narrow (10\arcsec\ on a side) and wide  (40\arcsec\ on a side) field-of-view modes. The LGS provided the wavefront reference source for AO correction, with the exception of tip-tilt motion.  Tip-tilt aberrations and quasi-static changes in the image of the LGS as seen by the wavefront sensor were measured contemporaneously with a second, lower-bandwidth wavefront sensor monitoring monitoring a nearby ($35\arcsec$) faint star (USNO-B1.0~0932-0296004, $R=16.9$~mag; \citealt{Monet2003}).

At each epoch, \simpbint\ was imaged in a variety of near-IR filters. We used the $J$ (1.25~\micron), $H$ (1.64~\micron), and $K_{\rm s}$ (2.15~\micron) broadband filters. We also used the CH$_{4}$S medium-band filter, centered at 1.592\,\micron\ (bandwidth of 0.126~\micron) around the $H$-band flux typical of T dwarfs.  For each filter, a series of dithered images was obtained, offsetting the telescope by a few arc-seconds between every 1--2 images. The sodium laser beam was pointed at the center of the NIRC2 field-of-view for all observations.

The images were reduced in a standard fashion, first by constructing a flat field from the differences of images of the telescope dome interior, with and without continuum lamp illumination.  A master sky frame was then created from the median of the bias-subtracted, flat-fielded images which was subtracted from individual frames. The results described here were based on the analysis of individual images and uncertainties were derived from the dispersion of individual measurements. 

An analytic model of the point spread function (PSF) as the sum of three elliptical Gaussians was used for estimating the flux ratio and relative positions of {\simpbint}'s components. The flux ratio, separation, and position angle of the binary were fitted to individual images in all filters. The averages of the results were adopted as the final measurements and the standard deviations as the errors. The relative astrometry was corrected for instrumental optical distortion, based on analysis by B. Cameron (priv.\ comm.) of images of a precisely machined pinhole grid located at the first focal plane of NIRC2. After applying this distortion correction, the 1$\sigma$ residuals of the pinhole images are at the 0.6\,mas level over the detector field of view. Since the binary separation and the dither steps are relatively small, the effect of the distortion correction is minor.

To convert the instrumental measurements of the binary's separation and PA into celestial units, we used a weighted average of the calibration from \citet{Pravdo2006}, with a pixel scale of $9.963\pm0.011$\,mas/pixel and an orientation for the detector's +y~axis of $+0.13\pm0.077^{\circ}$ east of north for NIRC2's narrow camera optics \citep{Ghez2008}, and $39.864\pm0.018$\,mas/pixel and $-0.16\pm0.07^{\circ}$ for NIRC2's wide camera optics. These values agree well with Keck Observatory's notional calibrations of $9.942\pm0.05$\,mas/pixel (narrow camera), $0.039686\pm0.05$~mas/pixel (wide camera), and $0.0\pm0.57^{\circ}$, as well as the $9.961\pm0.007$~mas/pixel and $-0.015\pm0.1347^{\circ}$ reported by \citet{Konopacky2007} for the narrow camera and $39.82\pm0.25$\,mas/pixel and $1.24\pm0.107^{\circ}$ from \citet{Metchev2004} and \citet{Metchev2005}.

A similar analysis was performed for \simpbinl; the uncertainties on separation are dominated by the 1\% plate-scale uncertainty ($\sim$$10$\,mas) and not by the uncertainties derived from the image-to-image dispersion ($\sim$$2$\,mas). The astrometric and photometric measurements for \simpbinl\ and \simpbint\ are reported in Tables~\ref{tbl-1} and \ref{tbl-2}, respectively.  Reassuringly, the $J$, $H$ and $K_{\rm s}$ contrast ratios measured for \simpbint\ obtained with NIRC2 (LGS AO) and NIRI (seeing limited) are in full agreement despite the very different observational setups.


\section{Spectral typing}
\label{typing}

We derived spectral types for {\simpbinl}A and {\simpbinl}B (respectively L$4.5\pm0.5$ and L$5.5\pm1.0$) using the \citet{Geballe2002} and \citet{Reid2001} spectral indices. For {\simpbint}A and  {\simpbint}B (respectively T$2.5\pm0.5$ and T$4.0\pm0.4$) we used the \citet{Burgasser2006} spectral indices. For the former binary, we chose not to use the more recent \citet{McLean2003} indices as they are defined over narrower wavelength intervals and provide little constraint on the spectral type for a noisy spectrum such as that of {\simpbinl}B. In all cases, the spectral type was taken to be the mean value derived from all the indices listed in Tables \ref{tbl-1} and \ref{tbl-2}. For each object, the uncertainty on its spectral type is defined as the dispersion of the measured indices. The spectral types were also confirmed visually by comparing the observed spectra with these of standard objects of similar types. The template spectra are from \citet{Knapp2004} and \citet{Chiu2006}\footnote{See: http://staff.gemini.edu/$\sim$sleggett/LTdata.html }. The {\simpbint}A and {\simpbint}B spectra closely match that of SDSS0851+18 (L4.5). The unresolved, far-red, photometry of the {\simpbinl} system by SDSS and MegaCam is consistent with a mid-L spectral type \citep{Chiu2006}, but provides no further constraint on the nature of this binary. The $H$-band spectrum of {\simpbint}B is nearly identical to that of 2MASSJ\,2254+31 (T4) and its $J$- and $K$-band spectra also closely match those of this T4 after re-normalization of each band. {\simpbint}A has shallower CH$_4$ absorption in both $H$ and $K$ bands compared to 2MASS1209-10, confirming its slightly earlier spectral type. The spectroscopic indices for both {\simpbint}A and {\simpbint}B are representative of objects of their spectral class. The only spectroscopic index where they fall outside of the bulk of the distribution for early Ts is the K/J index (similar to the $J-${\Ks} color).

\section{Discussion}
\label{discussion}
Both {\simpbint}  and {\simpbinl}  join the short list of L and T dwarfs binaries amenable to spatially resolved observation without the aid of adaptive optics. This is particularly important as $0.6-1.0$\,$\mu$m spectroscopy and photometry are useful for characterizing brown dwarfs and AO systems cannot, yet, reliably work blueward of $\sim1$\,$\mu$m. There are only 7 other known brown dwarf binaries with an apparent separation above $\sim0.5\arcsec$:  Gl 337CD ($0.53\arcsec$, L8+L8/T, no resolved spectroscopy), 2MASS J09153413+0422045AB ($0.73\arcsec$, L7+L7, no resolved spectroscopy published but flux ratio of components in $J$, $H$ and {\Ks} close to unity), 2MASS J15200224-4422419AB ($1.174\arcsec$, L1.5+L4.5, resolved spectroscopy by  \citealt{Burgasser2007b}), 2MASS J17072343-0558249Bab ($0.99\arcsec$, M9/L0+L3, resolved spectroscopy by \citealt{McElwain2006}),  DENIS-P J220002.05-303832.9B (1.094\arcsec, M9+L0, resolved spectroscopy by \citealt{Burgasser2006d}), $\epsilon$ Indi Bab ($0.732\arcsec$, T1.5+T6, resolved spectroscopy by \citealt{Kasper2009, King2010}) and SDSS1416+1348AB ($9\arcsec$, L7+T7.5, resolved spectroscopy by \citealt{Burningham2010}).

\begin{figure*}[!htb]
\epsscale{.8}
\plotone{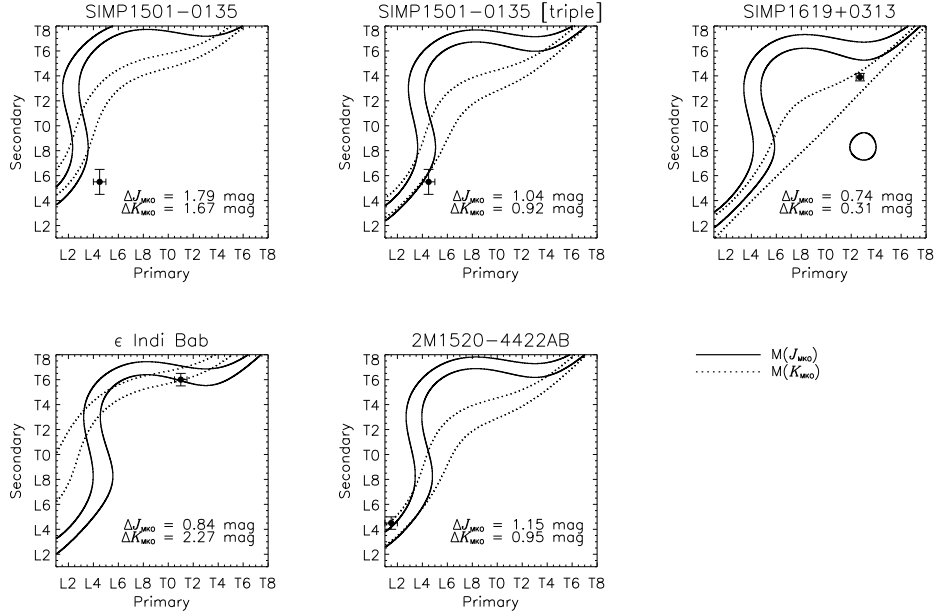}
\caption{\label{fig5}  Primary versus secondary spectral type for given values of $\Delta J$ and $\Delta${\Ks}. The region constrained by the $J$-band and $K_{\rm s}$-band flux ratios are given by the continuous curve and dotted curves, respectively. The width of the regions account for the dispersion in the spectral type versus absolute magnitude relation. The \citet{Liu2006} M$_J$ and M$_{\Ks}$ versus spectral type polynomial relations were used to determine the allowed primary versus secondary spectral type for a given flux ratio. The upper panels illustrate the cases of {\simpbinl} in the binary scenario, in the triple scenario and \simpbint. The lower panels illustrates the cases of two previously known resolved brown dwarf pairs, $\epsilon$ Indi Bab and 2M1520-4422AB. The circle in the \simpbint\ panel arises from the ambiguity around the $J$-band brightening in the spectral type versus absolute magnitude relation. A companion $0.74$\,mag fainter in $J$ than a T2.5 could either be a L7-L9 or a T5-T6 dwarf.}
\end{figure*}

In principle, the spectral types of both components of a binary can be deduced from their $\Delta J$ and $\Delta K_{\rm s}$ flux ratio using predetermined $M_J$ and $M_{K{\rm s}}$ versus spectral type relations (e.g.: \citealt{Liu2006}). The flux ratios in each photometric band define a narrow region of allowed primary versus secondary spectral types, and any binary falling outside of the intersect of these regions would have peculiar photometric properties worthy of further investigation. We went through this exercise for {\simpbinl}AB, {\simpbint}AB, $\epsilon$ Indi Bab and  2M1520-4422AB and show the results on Figure~\ref{fig5}. We used the mean of the magnitude versus spectral type relations with and without binaries given by \citet{Liu2006} as this is in very good agreement with parallax measurements on a large sample of single L and T dwarfs (Liu, personal communication)\footnote{This gives absolute magnitudes fainter, by up to 0.3\,mag, than the \citet{Knapp2004} polynomial relations.}. Note that for early and mid-T dwarfs, as is the case for {\simpbint}, a fainter secondary in the $J$-band could in principle have an earlier spectral type than its primary. As visible in Figure~\ref{fig5}, neither  {\simpbinl} nor  {\simpbint} fall within the intersect region defined by this diagram; both systems are challenging to explain with standard models. 

\begin{figure*}[!htb]
  \epsscale{.8}
\plotone{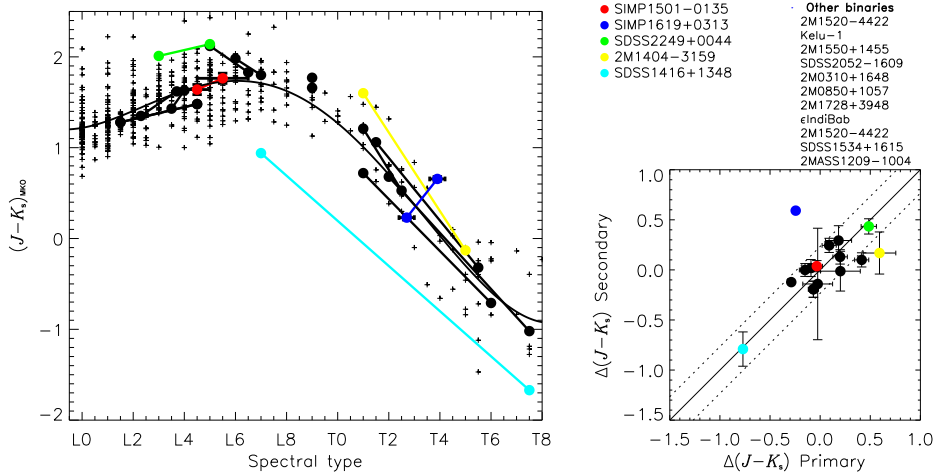}
\caption{\label{fig2} Left panel : color versus spectral type for all L and T dwarfs with a near-infrared spectral type and $J-${\Ks} photometric accuracy better than 0.25\,mag. Components of binaries with published resolved spectral types are shown linked by a straight line. The continuous line spanning the diagram gives the 4th order polynomial fit to the data. \simpbint, {\simpbinl} and four binaries that are redder or bluer than the bulk of other objects of similar spectral types are shown with color symbols for clarity. Right panel: Difference between the $J-${\Ks} of each component of binaries and the fit to the field L and T dwarf $J-${\Ks} versus spectral type relation. The straight line shows a one-to-one relation between the $\Delta$($J-${\Ks}) of the primary and secondary while the dotted lines show the dispersion ($\pm1\sigma$) of field objects from the polynomial relation.  Point without error bars (e.g.: \simpbint) have photometric uncertainties smaller than 0.05\,mag. Most binaries either have two components redder or bluer than the bulk of the distribution, with {\simpbint} being the only significant exception. Photometry and spectral types for field binaries are from \citet{Liu2006, Liu2010, Allers2010, Burgasser2007b, Liu2005, Looper2008, Burningham2010, Stumpf2010, Burgasser2010_binaires}.}
\end{figure*}

\subsection{{\simpbinl}, a triple L dwarf?}

For \simpbinl, the L4.5 primary and $J$-band flux ratio of 1.79\,mag would imply a late-T secondary based on $M_J$ versus spectral type relations. This can readily be discarded from the available spectroscopy. The strong $H$-band CH$_4$ absorption, the hallmark of T dwarfs, would readily be detected in the resolved NIRI spectroscopy, but it is absent. Assuming that {\simpbinl}A is itself an unresolved equal-luminosity binary reduces the flux ratio of either component of {\simpbinl}A to {\simpbinl}B by 0.75\,mag and moves the system to the edge of the allowed region in the diagram shown in Figure~\ref{fig5}. This scenario has not been confirmed by AO observation. Using {\simpbinl}B as a reference PSF, we found no significant difference between the components of {\simpbinl}, which puts an upper limit of $0.05\arcsec$ on the separation for a near-equal luminosity inner binary. The ratio of periods in triple systems with components of similar masses is limited by stability constraints. In stellar systems, this ratio is never less than 5 \citep{Tokovinin2006}, which would correspond to a maximum separation of $\sim0.38\arcsec$ here. By putting an upper limit to a separation of  $\sim0.05\arcsec$, using LGS AO observation, allowed us to test period ratios ranging from $5$ to $\sim100$. We nevertheless argue that it is the most simple explanation for the odd properties of this system. One other peculiarity of the {\simpbinl} system is that it forms one of the widest known field brown dwarf binary, second to the L6+T7.5 system SDSS J1416+1348AB \citep{Burningham2010}. If the system is indeed a triple, the inner pair, as yet unresolved, could have dynamically excited the outer companion. 

 High resolution spectroscopy could help to test this hypothesis by verifying if {\simpbinl}A is a double-lined binary. Higher resolution imagery, for example AO on a 30\,m-class telescope, could also be an observational mean of testing this hypothesis. 

The {\simpbinl} system is similar to the recently reported late-L binary 2MASS J0850+1057 \citep{Burgasser2010_binaires}. The resolved photometry of this latter system suggests that the components are L7 for the brighter ($\Delta J=1.13$\,mag) primary and L6 for the secondary. The possibility that the brightness reversal of this object (brighter later-type component) is due to an unresolved primary is explored by \citet{Burgasser2010_binaires} and could be tested through long-term orbital monitoring of the resolved pair or radial velocity. 

\subsection{{\simpbint} and known field binaries}

The spectral type versus $J-${\Ks} diagram for L and T dwarfs shows a mild ($\sim0.5$\,mag) reddening of objects through the L dwarf sequence, followed by a progressively bluer $J-${\Ks} color through the T dwarf sequence (see Figure \ref{fig2}). This overall behavior is understood in terms of the evolution of dust grains through the L dwarfs and their subsequent settling below the photosphere of T dwarfs. However, for any given spectral type, a significant scatter ($\sim0.3$\,mag) remains that is unaccounted for by the observational uncertainties. This scatter is thought to largely arise from differences in surface gravity, metallicity and dust clouds properties: low gravity and high metallicity accounting for the reddest objects. In that context, one would expect the two components of a binary to roughly lie at the same location relative to the distribution of field objects of similar types (see right panel of Figure \ref{fig2}). For example, both components of the L7+T7.5 binary SDSS1416+1348 \citep{Burningham2010, Bowler2010, Schmidt2010} are significantly bluer than field late-Ls and late-Ts, which could mean the system is relatively old and thus has a high gravity and a low metallicity. Other binaries such as SDSS2249+0044, and to a lesser extent 2M1404-3159  \citep{Looper2008, Allers2010}, have components that are noticeably redder than field objects and likely to be relatively young and/or metal rich. The \simpbint\ components are remarkable as they straddle the distribution, {\simpbint}A being bluer than similar-type objects while {\simpbint}B being redder.

Figure~\ref{fig5} shows that {\simpbint} falls within the region defined by its {\Ks} flux ratio but it is significantly off from what is expected from its $J$ flux ratio. This is also clearly seen in the spectroscopy shown in Figure\,\ref{fig3}. {\simpbint}A closely matches the reference T3, 2MASS1209-10, over the whole near-infrared domain. Between 1.4\,$\mu$m and 2.45\,$\mu$m, {\simpbint}B is basically identical to 2MASS\,J2254+31, but it is about 0.8\,mag brighter in the 1.0-1.3\,$\mu$m interval, as expected from its very blue $J-H$ color.

The right panel of Figure~\ref{fig2} shows the difference between the $J-K_{\rm s}$ color of both components of all published binaries that have resolved $J$ and {\Ks} to the $J-K_{\rm s}$ for mean field dwarfs. The mean color of field dwarfs as a function of spectral type was calculated from a fourth order polynomial fit to the field L and T dwarfs plotted in Figure~\ref{fig2}: $J-K_{\rm s}({\rm MKO}) = 1.218+4.95\times10^{-2} \times {\rm spt} + 3.21\times10^{-2}  {\rm spt}^2 -5.33\times10^{-3}  {\rm spt}^3 + 1.69\times10^{-4}  {\rm spt}^4$, where L0 corresponds to ${\rm spt}=0$ and T8 to ${\rm spt}=18$. The figure shows a clear one-to-one correlation between the $J-K_{\rm s}$ color offsets of the primary and secondary components. All published L and T binaries fall within $\pm$\,$1$\,$\sigma$ of the scatter (2M1404-3159 does so only marginally). {\simpbint} is the only known outlier to this distribution, with {\simpbint}A being 0.25\,mag redder than a typical T2.5 and {\simpbint}B being 0.6\,mag bluer than a typical T4.0 color.

\begin{figure}[!htb]
 \plotone{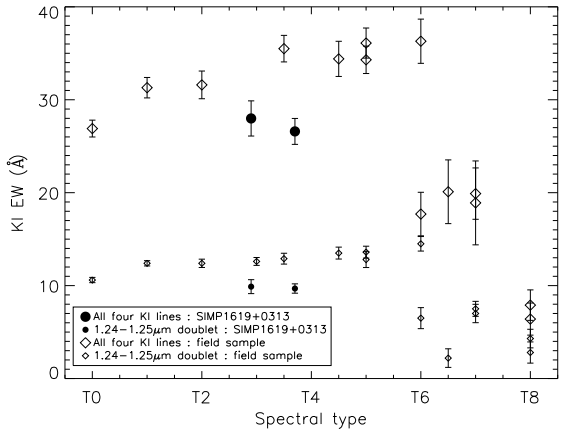}
 \caption{\label{fig7} Equivalent widths of the two KI doublets and the $1.24-1.25\,\mu$m doublet only for {\simpbint}A, {\simpbint}B and field T dwarfs from the \citet{McLean2003} sample.}
 \end{figure}

One noteworthy point from the {\simpbint} spectroscopy is that the $J$-band KI doublets of both components are significantly narrower than for field T dwarfs (i.e. the \citet{McLean2003} sample; see Figure \ref{fig7}). A smaller equivalent width of the KI doublet is generally related to low metallicity \citep{Bowler2009} and/or low surface gravity \citep{McGovern2004}. It is unclear how this property relates, if it does, to the peculiar photometric properties of the system.

\subsection{SIMP1619+0313AB: a testbed for early-T cloud models? \label{section_interpretation}}
We propose that the photometric properties of \simpbint\ can be explained if the atmosphere of both components is described as a superposition of regions with and without dust-bearing clouds, with very different fractional coverage for {\simpbint}A and {\simpbint}B. This idea has been proposed to explain the photometric properties of L/T transition objects \citep{Marley2010, Marley2003} and the photometric variability of a T2.5 dwarf \citep{Artigau2009}. 

As a toy model, we supposed that the emerging flux of each component was the weighted sum of the flux from two types of atmosphere: dust settling or dust-free (respectively {\it cloudy}  and {\it clear} by \citealt{Burrows2006}).

We caution that this description must be seen as a toy model only; a complete physical description is beyond the scope of this paper. Clouds directly impact the pressure-temperature profile of the atmosphere compared to cloud-free models, and only a superposition of two types of regions with the same underlying structure can accurately represent a real atmosphere. Recent models by \citet{Marley2010} address this question in an attempt to explain the L/T transition self-consistently. {\simpbint} would be an interesting test case for such patchy cloud models.

The properties of the pair can be reproduced to first order if {\simpbint}A is covered by similar fractions of {\it clear} and {\it cloudy} (respectively $44\pm2$\% and $56\pm2$\%) atmosphere while {\simpbint}B has a similar temperature with only about $7.3\pm1.4$\% of its surface covered by {\it clear} regions. The derived uncertainties in the filling factors only account for the effects of the accuracy of near-infrared photometry and do not include uncertainties in the models. For both objects we assumed that the {\it clear} phase was 100\,K warmer than the {\it cloudy} phase. Determining the exact temperature difference depends on the temperature profile of the atmosphere and requires a detailed physical simulation. A $\sim100$\,K temperature difference between these two phases has been derived for an early-T dwarf displaying photometric variability \citep{Artigau2009} .

\begin{figure}[!htb]
\epsscale{0.75}
\plotone{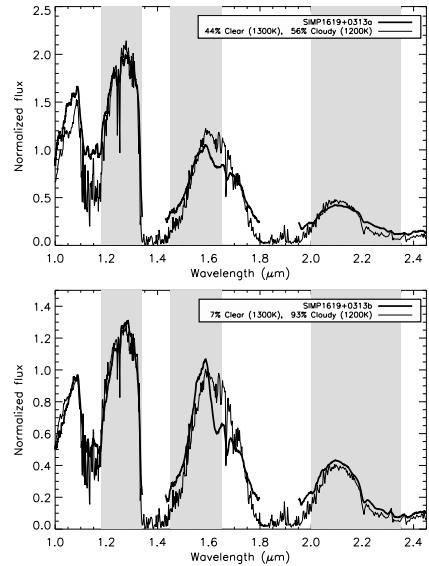}
\caption{\label{fig6} {\simpbint}A (upper panel) and {\simpbint}B (lower panel) with the best fit assuming a superposition of  {\it cloudy} and {\it clear} model atmosphere \citep{Burrows2006}  on the surface of each brown dwarf. The shaded regions represents the wavelength intervals over which the models were adjusted to the observed spectrum. The fit avoided deep telluric absorption and the $H$-band CH$_4$ absorption.}
\end{figure}

Figure\,\ref{fig6} shows the best fit superposition of {\it clear} and {\it cloudy} spectra for both components. These fits predict a $K$-band flux difference of 0.06\,mag between these two components, compared to the observed $\Delta K_s=0.31$\,mag. The lower contrast in the toy model may point toward a slightly larger temperature difference between {\it clear} and {\it cloudy} regions. It is noteworthy that the derived effective temperature difference between both objects determined from the simplistic model fit done here (T=$1247$\,K and T$=1208$\,K, for $\Delta$T$=39$\,K) is very close to the temperature difference found through the \citet{Stephens2009} polynomial fit ($\Delta$T$=55$\,K, See Table\,\ref{tbl-1}).

\subsection{Far-red colors of SIMP1619+0313AB  \label{section_izj_simp1619}}
The disappearance of dust in the atmosphere of a T dwarf increases the relative importance of the pressure-broadened KI doublet at $\sim0.77$\,$\mu$m, leading to redder $i-z$ and $z-J$ colors for dust-free objects. The far-red colors of {\simpbint}AB therefore contain additional constraints to the toy model proposed here, wherein variation of the dust content is the key to explain the odd near-infrared color inversion of this binary.

An interesting result from the far-red photometry is the difference in color $i-z$ and $z-J$ of the two components; in both colors, the warmer {\simpbint}A is redder ($i-z=3.37$, $z-J=4.08$) than the cooler {\simpbint}B ($i-z=3.26$, $z-J=3.76$). This is opposite to the overall trend of early-to-mid-T dwarfs which become redder with later spectral types. It is also opposite to what either {\it clear} or {\it cloudy} models alone would predict. The toy model proposed earlier provides a natural explanation for this color reversal in far-red colors; {\it cloudy} models are in both colors bluer than {\it clear} models 100\,K warmer. For the fractional coverage given earlier, a 1200\,K  {\it cloudy} phase and 1300\,K  {\it clear} phase, one would expect {\simpbint}A to be redder than {\simpbint}B in $z-J$ ($\sim 0.15$\,mag) and in $i-z$ ($\sim0.10$\,mag). These values are in rough agreement with the differences observed between the components, although in the toy model the two components are bluer in $i-z$ than observed by $\sim0.4\,$mag. This could be accounted for by a higher gravity ($\log g = 5.5$) for both object. However, it is not possible to fit  the near-infrared SED with such a high gravity. High surface gravity models are bluer in $J-K$ due to an increase in collision-induced absorption by molecular hydrogen in $K$ and cannot reproduce objects as red as {\simpbint}B.

Lower metallicity, as suggested by the low KI doublet equivalent widths observed for both components of the system (See Figure\,\ref{fig7}) further increases the $i-z$ and $z-J$ color discrepancy. Models with Z=0.3 are bluer in $z-J$ by $\sim0.2$\,mag compared to solar metallicity ones.

\section{Conclusions and Summary}
We reported the discovery and characterization of two widely separated brown dwarf binaries: {\simpbinl} and {\simpbint}. Based on photometric measurements, {\simpbinl} is most likely to be a triple system, with {\simpbinl}A being a near-equal luminosity mid-L dwarf and {\simpbinl}B a slightly later mid-L. This system has not been resolved as a triple in AO observations. This could be due to small physical projected separation at the time of observations and hope remains that future observations may resolve the suspected inner binary. It is also be possible that the semi-major axis is too small ($<$2\,AU) to be resolved with current systems. In that case the system's binarity could be tested through very-high resolution spectroscopy to detect double lines or through still higher spatial resolution observation with either non-redundant mask imaging on 10-m class telescopes or future larger observatories. {\simpbinl} is expected to have a $\textgreater 475$\,yr orbital period (total mass $\textless0.12$M$_\odot$). If {\simpbinl}A proves to be an unresolved binary, then the expected orbital period for the outer component will be even larger; $\textgreater 750$\,yr for a total system mass of $\textless0.18$M$_\odot$.

The \simpbint\ components also display odd relative photometric properties: a near-infrared color reversal unlike anything seen in known L or T binaries. The toy model we presented shows that this inversion can be explained by the two components of the binary having very different dust content. This scenario may be tested with a more complete characterization of the SED through mid-infrared imaging and far-red spectroscopy. The relatively wide separation of the system implies that under good observing conditions (i.e. seeing better than about $0.5\arcsec$), resolved spectroscopy can be obtained without the aid of adaptive optics, simplifying follow-up work. {\simpbint} is expected to have a $\textgreater 175$\,yr period (total mass $\textless0.12$M$_\odot$). Neither system will provide significant constraints on dynamical masses within a useful time baseline.

\begin{table}[!tbp]
{\tiny
\begin{center}
\caption{Observations log\label{tbl-3}}
\begin{tabular}{lcccc}
\hline\hline
Instrument & Telescope & date          & filter & integration \\
\hline
\multicolumn{5}{c}{\simpbinl}\\
\hline
NIRI$^{\rm a,b}$     & Gemini North& 2008/04/28 & $J^{\rm g}$  & 100\,s\\
NIRI$^{\rm a,b}$     & Gemini North& 2008/04/28 & $H^{\rm h}$  & 100\,s\\
NIRI$^{\rm a,b}$     & Gemini North& 2008/04/28 & $K^{\rm i}$  & 100\,s\\
NIRI$^{\rm a,c}$     & Gemini North& 2008/05/19 & $J$  & 1200\,s\\
NIRI$^{\rm a,c}$     & Gemini North& 2008/05/19 & $H$  & 1200\,s\\
NIRI$^{\rm a,c}$     & Gemini North& 2008/05/19 & $K$  & 1200\,s\\
NIRI LGS$^{\rm a,d}$ & Gemini North& 2008/05/23 & $J$  & 810\,s\\
NIRI LGS$^{\rm a,d}$ & Gemini North& 2008/05/23 & $H$  & 810\,s\\
NIRI LGS$^{\rm a,d}$ & Gemini North& 2008/05/24 & $K_{\rm s}$& 810\,s\\
CPAPIR$^{\rm e}$     &OMM& 2006/04/11 & $J$  &  24\,s\\ 
CPAPIR$^{\rm e}$     &OMM& 2009/02/17 & $J$  &  1201\,s\\ 
CPAPIR$^{\rm e}$     &OMM& 2009/02/17 & $H$  & 1169\,s\\ 
CPAPIR$^{\rm e}$     &OMM& 2009/03/03 & $K_{\rm s}$&   568\,s\\ 
MegaCam$^{\rm e}$ &CFHT& 2006/07/20 & $i$ & 200\,s\\
\hline
\multicolumn{5}{c}{\simpbint}\\
\hline
NIRI$^{\rm a,b}$     & Gemini North& 2008/05/18 & $J^{\rm g}$  &  900\,s\\
NIRI$^{\rm a,b}$     & Gemini North& 2008/05/18 & $H^{\rm h}$   & 1170\,s\\
NIRI$^{\rm a,b}$     & Gemini North& 2008/05/18 & $K^{\rm i}$   &  900\,s\\
NIRI$^{\rm a,c}$     & Gemini North& 2008/05/03 & $J$  &  100\,s\\
NIRI$^{\rm a,c}$     & Gemini North& 2008/05/03 & $H$  &  100\,s\\
NIRI$^{\rm a,c}$     & Gemini North& 2008/05/03 & {\Ks}  &  100\,s\\
NIRC2$^{\rm d}$      &Keck-II& 2008/04/01 & $J$  &      540\,s     \\
NIRC2$^{\rm d}$      &Keck-II& 2008/04/01 & $H$  &    540\,s      \\
NIRC2$^{\rm d}$      &Keck-II& 2008/04/01 & {\Ks}&    540\,s       \\
NIRC2$^{\rm d}$      &Keck-II& 2008/05/29 & CH$_4$s  &    200\,s      \\
NIRC2$^{\rm d}$      &Keck-II& 2008/05/29 & {\Ks}&    360\,s       \\
CPAPIR$^{\rm e}$     &OMM& 2006/08/21 & $J$  &  24\,s\\ 
CPAPIR$^{\rm e}$     &OMM& 2009/03/04 & $H$  &   601\,s\\ 
CPAPIR$^{\rm e}$    &OMM & 2009/03/04 & $J$  &  1039\,s\\  
CPAPIR$^{\rm e}$     &OMM& 2009/03/09 & {\Ks}&  1039\,s\\  
MegaCam$^{\rm f}$&CFHT & 2006/08/29 & $i$ &  200\,s \\
MegaCam$^{\rm f}$ &CFHT& 2009/06/01 & $z$ &  100\,s \\
\hline\hline
\end{tabular}
\tablenotetext{a}{Gemini program ID : GN-2008A-Q-57.}
\tablenotetext{b}{Seeing limited, resolved spectroscopy.}
\tablenotetext{c}{Seeing limited, resolved photometry.}
\tablenotetext{d}{LGS-AO imaging.}
\tablenotetext{e}{Unresolved astrometry and photometry.}
\tablenotetext{f}{Resolved photometry.}
\tablenotetext{g}{NIRI ``$J$'' spectroscopy setup, useful domain : $0.985\mu$m - $1.35\mu$m.}
\tablenotetext{h}{NIRI ``$H$'' spectroscopy setup, useful domain : $1.36\mu$m - $1.89\mu$m.}
\tablenotetext{i}{NIRI ``$K$'' spectroscopy setup, useful domain : $1.90\mu$m - $2.49\mu$m.}

\end{center}
}
\end{table}

\begin{table}[!htb]
{\tiny
\begin{center}

\caption{Parameters of {\simpbint} \label{tbl-1}}
\begin{tabular}{lccc}
\hline\hline 
Parameter& AB & A & B\\
\hline
2MASS designation$^{\rm a}$&J16192751+0313507 & & \\
$\alpha^{\rm a}$&$244.864664$&&\\
$\delta^{\rm a}$&$+03.230753$&&\\

$i[AB]_{\rm MegaCam}$ & $22.50\pm0.09$&$23.11\pm0.14$&$23.41\pm0.19$\\

$z[AB]_{\rm MegaCam}$ & $19.17\pm0.03$&$19.74\pm0.05$&$20.15\pm0.20$\\

$J^{\rm a} _{\rm 2MASS}$&$15.454\pm0.071$ & &\\
$J_{\rm 2MASS}^{\rm b}$  &  $15.415\pm0.009$ & $15.851\pm0.013$ & $16.616\pm0.013$\\
$J_{\rm MKO}^{\rm b}$    &  $15.207\pm0.008$ & $15.652\pm0.012$ & $16.390\pm0.012$\\

$H^{\rm a} _{\rm 2MASS}$&$15.022\pm0.098$&&\\
$H_{\rm 2MASS}^{\rm b}$  &  $14.916\pm0.012$ & $15.482\pm0.015$ & $15.894\pm0.015$\\
$H_{\rm MKO}$    &  $14.974\pm0.012$ & $15.543\pm0.015$ & $15.948\pm0.015$\\

{\Ks}$^{\rm a}_{\rm 2MASS}$&$15.015\pm0.123$&&\\
$K_{\rm 2MASS}^{\rm b}$  &  $14.925\pm0.012$ & $15.493\pm0.015$ & $15.901\pm0.015$\\
$K_{\rm MKO}^{\rm b}$    &  $14.868\pm0.012$ & $15.432\pm0.015$ & $15.848\pm0.015$\\

$\Delta J$ NIRC2&$0.738\pm0.009$&&\\
$\Delta J$ NIRI &$0.718\pm0.008$&&\\

$\Delta H$ NIRC2&$0.405\pm0.008$&&\\
$\Delta H$ NIRI &$0.388\pm0.012$&&\\

$\Delta {\Ks}$ NIRC2&$0.312\pm0.015$&&\\
$\Delta {\Ks}$ NIRI &$0.264\pm0.005$&&\\

Separation&$0.691\pm0.002\arcsec$&&\\
Position angle&$71.23\pm0.23\,^\circ$&&\\
$\mu_\alpha \cos\delta$&$70\pm7$\,mas/yr&&\\
$\mu_\delta$&$-289\pm9$\,mas/yr&&\\
H2O-J$^{\rm c}$&&$0.510$ $[\rm{T}2.4]$&$0.415$ $[\rm{T}3.5]$\\
CH4-J$^{\rm c}$&&$0.612$ $[\rm{T}2.7]$&$0.542$ $[\rm{T}3.6]$\\
H2O-H$^{\rm c}$&&$0.502$ $[\rm{T}2.4]$&$0.415$ $[\rm{T}4.3]$\\
CH4-H$^{\rm c}$&&$0.786$ $[\rm{T}3.1]$&$0.590$ $[\rm{T}4.1]$\\
CH4-K$^{\rm c}$&&$0.532$ $[\rm{T}3.1]$&$0.374$ $[\rm{T}3.9]$\\
K/J $^{\rm c}$&&$0.172$&$0.182$\\
Near-IR SpT&&$\rm{T}2.5\pm0.5$&$\rm{T}4.0\pm0.5$\\
Photometric distance$^{\rm d}$&$22\pm3$\,pc&$22\pm3$\,pc&$22\pm3$\,pc\\
Physical separation$^{\rm d}$&$15.4\pm2.1$\,AU&&\\
Temperature$^{\rm e}$&&$1219\pm100$\,K&$1164\pm100$\,K\\
\hline\hline

\end{tabular}
\tablenotetext{a}{From the 2MASS point source catalogue \citep{Skrutskie2006}.}
\tablenotetext{b}{Combining the unresolved CPAPIR photometry with the NIRC2 flux ratios.}
\tablenotetext{c}{Index as defined by \citet{Burgasser2006}.}
\tablenotetext{d}{Using the M$_{K_{\rm s}}$ versus spectral type relation by \citet{Liu2006}. Considering the peculiar colors of $J$-band flux ratio, we preferred M$_{\Ks}$ to the M$_J$ polynomial relation for distance estimation. Uncertainty in the photometric distance include the {\Ks}-band photometric uncertainty and an assumed 0.2\,mag scatter in the M$_{\Ks}$ versus spectral type relation.}
\tablenotetext{e}{Using \citet{Stephens2009} polynomials; assumes spectral types of T2.7 and T3.9 (i.e. arithmetic means of spectral types derived from individual indices).}

\end{center}
}
\end{table}

\begin{table}[!htb]
{\tiny
\begin{center}
\caption{Parameters of \simpbinl\label{tbl-2}}
\begin{tabular}{lccc}
\hline\hline
Parameter &AB& A & B\\
\hline	
2MASS designation$^{\rm a}$&J15015302-0135068&&\\
SDSS designation&J150153.00-013507.2&&\\

$\alpha$$^{\rm a}$&225.470943&&\\
$\delta$$^{\rm a}$&-01.585236&&\\

$r_{\rm SDSS}$&$22.706\pm0.248$&&\\
$i_{\rm SDSS}$&$20.802\pm0.071$&&\\
$z_{\rm SDSS}$&$18.622\pm0.048$&&\\

$i[AB]_{\rm MegaCam}$ & $20.44\pm0.03$&&\\

$J^{\rm a}_{\rm 2MASS}$&$16.081\pm0.094$&&\\
$J_{\rm 2MASS}^{\rm b}$  &  $15.978\pm0.027$ & $16.168\pm0.045$ & $17.960\pm0.045$\\
$J_{\rm MKO}^{\rm b}$    &  $15.843\pm0.027$ & $16.033\pm0.045$ & $17.827\pm0.045$\\

$H^{\rm a} _{\rm 2MASS}$&$14.952\pm0.071$&&\\
$H_{\rm 2MASS}^{\rm b}$  &  $14.895\pm0.015$ & $15.101\pm0.019$ & $16.800\pm0.019$\\
$H_{\rm MKO}^{\rm b}$    &  $14.956\pm0.015$ & $15.163\pm0.019$ & $16.858\pm0.019$\\

{\Ks}$^{\rm a}_{\rm 2MASS}$&$14.257\pm0.086$&&\\
$K_{2MASS}^{\rm b}$  &  $14.122\pm0.028$ & $14.332\pm0.034$ & $16.010\pm0.034$\\
$K_{MKO}^{\rm b}$    &  $14.182\pm0.027$ & $14.393\pm0.033$ & $16.063\pm0.033$\\

$\Delta J_{NIRI}$&$1.79\pm0.04$&&\\
$\Delta J_{NIRI LGS}$ &$1.77\pm0.05$&&\\

$\Delta H_{NIRI}$ &$1.70\pm0.01$&&\\
$\Delta H_{NIRI LGS}$&$1.67\pm0.05$&&\\

$\Delta K_{s NIRI}$&$1.67\pm0.02$&&\\
$\Delta K_{s NIRI LGS}$ &$1.62\pm0.05$&&\\

Separation&$0.96\pm0.01\arcsec$&&\\
Position angle&$331.4\pm0.1\,^\circ$&&\\

$\mu_\alpha \cos\delta$&$-225\pm10$\,mas\,yr$^{-1}$&&\\
$\mu_\delta$&$-69\pm12$\,mas\,yr$^{-1}$&&\\

H$_2$O$^A$  $^{\rm c}$   &&$0.582$[L4.5]&$0.541$[L6.0]\\
H$_2$O$^B$  $^{\rm c}$   &&$0.673$[L4.0]&$0.672$[L4.0]\\
H$_2$O$^C$  $^{\rm c}$   &&$1.697$      &$1.788$\\
H$_2$O$^D$  $^{\rm c}$   &&$0.673$      &$0.706$\\

H2O 1.5$\mu$m$^{\rm d}$   &&$1.520$[L4.5]&$1.482$[L3.5]\\
CH$_4$ 2.2$\mu$m$^{\rm d}$&&$0.987$[L5.0]&$1.114$[L8.0]\\

Near-IR SpT&&L$4.5\pm0.5$&L$5.5\pm1.0$\\
Photometric distance$^{\rm e}$ && $31\pm5$ / $44\pm7$\,pc$^{\rm f}$  &  $56\pm8$\,pc\\
Physical separation&$30\pm5$\,AU / $47\pm7$\,AU$^{\rm g}$&&\\
\hline\hline

\end{tabular}
\tablenotetext{a}{From the 2MASS point source catalogue \citep{Skrutskie2006}. }
\tablenotetext{b}{Combining the unresolved CPAPIR photometry with the NIRI flux ratios. }
\tablenotetext{c}{Index as defined by \citet{Reid2001}.}
\tablenotetext{d}{Index as defined by \citet{Geballe2002}.}

\tablenotetext{e}{Using the \citet{Liu2006} M$_{K_{\rm s}}$ versus spectral type relations.}
\tablenotetext{f}{Respectively assuming that {\simpbinl}A is itself a single object and an equal luminosity binary.}
\tablenotetext{g}{Assuming a $31\pm5$ and $49\pm7$\,pc distance respectively.}
\end{center}
}
\end{table}

\clearpage

\acknowledgments
The authors would like to thank Marcel Bergman, the Gemini queue observer who took the {\simpbinl} and {\simpbint} observations, for notifying them that two of the SIMP brown dwarf candidates "looked like doubles" in GNIRS acquisition images. The authors would like to thank Mike Shara for his leadership and financial support in enabling the CPAPIR observing campaign on the 1.5-m CTIO telescope. Our special thanks to Alberto Pasten and Claudio Aguilera for carrying out the SIMP observations and to Philippe Vall\'ee for managing the SIMP data. This work was supported in part through grants from the Natural Sciences and Engineering Research Council, Canada, and from Fonds Qu\'eb\'ecois de la Recherche sur la Nature et les Technologies. CPAPIR was built through a grant from the Canada Foundation for Innovation. This research has benefited from the M, L, and T dwarf compendium housed at \hbox{DwarfArchives.org} and maintained by Chris Gelino, Davy Kirkpatrick, and Adam Burgasser. This publication makes use of data products from the Two Micron All Sky Survey, which is a joint project of the University of Massachusetts and the Infrared Processing and Analysis Center/California Institute of Technology, funded by the National Aeronautics and Space Administration and the National Science Foundation. Based on observations obtained at the Gemini Observatory (program ID : GS-2007A-Q-28 \& GN-2008A-Q-57), which is operated by the Association of Universities for Research in Astronomy, Inc., under a cooperative agreement with the NSF on behalf of the Gemini partnership: the National Science Foundation (United States), the Science and Technology Facilities Council (United Kingdom), the National Research Council (Canada), CONICYT (Chile), the Australian Research Council (Australia), CNPq (Brazil) and CONICET (Argentina). Some of the data presented herein were obtained at the W.M. Keck Observatory, which is operated as a scientific partnership among the California Institute of Technology, the University of California, and the National Aeronautics and Space Administration. The Observatory was made possible by the generous financial support of the W.M. Keck Foundation.

\bibliographystyle{apj}


\end{document}